\documentclass[a4paper]{jpconf}
\usepackage{graphicx}
\usepackage{amsmath,amssymb}
\begin{document}
\title{Anomaly in the cosmic-ray energy spectrum at GeV-TeV energies}

\author{Satyendra Thoudam$^{1,*}$ and J\"org R. H\"orandel$^{1,2}$}

\address{$^1$Department of Astrophysics, IMAPP, Radboud University Nijmegen, 6500 GL Nijmegen, The Netherlands}
\address{$^2$Nikhef, Science Park Amsterdam, 1098 XG Amsterdam, The Netherlands}

\ead{$^*$s.thoudam@astro.ru.nl}

\begin{abstract}
Recent measurements of cosmic rays by various experiments have found that the energy spectrum of cosmic rays is harder in the TeV region than at GeV energies. The origin of the spectral hardening is not clearly understood. In this paper, we discuss the possibility that the spectral hardening might be due to the effect of re-acceleration of cosmic rays by weak shocks associated with old supernova remnants in the Galaxy.
\end{abstract}

\section{Introduction}
The recent measurements of cosmic rays by the ATIC \cite{bib:Panov2007}, CREAM \cite{bib:Yoon2011}, and PAMELA \cite{bib:Adriani2011} experiments have revealed an anomaly in the cosmic-ray energy spectrum at GeV-TeV energies. The measured spectrum is found to be harder in the TeV region than at GeV energies. The spectral anomaly is difficult to explain using standard models of acceleration of cosmic rays and the nature of cosmic-ray propagation in the Galaxy, which predict a single power-law spectrum over the whole energy range.

Various explanations for the spectral anomaly have been proposed. These include hardening in the cosmic-ray source spectrum at high energies \cite{bib:Biermann2010, bib:Ohira2011, bib:Yuan2011, bib:Ptuskin2013}, changes in the cosmic-ray propagation properties in the Galaxy \cite{bib:Tomassetti2012, bib:Blasi2012}, and the effect of nearby sources \cite{bib:Thoudam2012, bib:Thoudam2013, bib:Erlykin2012, bib:Zatsepin2013}. In this paper, we discuss the possibility that the anomaly can be due to the effect of re-acceleration of cosmic rays by weak shocks during their propagation through the Galaxy \cite{bib:Ptuskin2011, bib:Thoudam2014}. The model assumes that cosmic rays are accelerated by strong supernova remnant shock waves. After acceleration, cosmic rays escape from the remnants and undergo diffusive propagation in the Galaxy. During the propagation, some fraction of the cosmic rays may again encounter expanding supernova remnant shock waves and get re-accelerated \cite{bib:Wandel1988, bib:Berezhko2003}. This encounter is expected to be more often with older remnants as they occupy a larger volume in the Galaxy as compared to the younger remnants. Therefore, this process of re-acceleration is expected to be produced mainly by weaker shocks, and as weaker shocks accelerate particles with a softer spectrum, the re-accelerated particles will have an energy spectrum which is steeper than the spectrum originally produced by the strong shocks. The re-accelerated component can dominate the GeV energy region, while the non-reaccelerated component (hereafter ``normal component") dominates at higher energies. In the following, we will show that this effect can explain the observed spectral anomaly.   

Re-acceleration can also be produced by the same magnetic turbulence responsible for the scattering and spatial diffusion of cosmic rays in the Galaxy. This process, commonly known as distributed re-acceleration, have been predicted to produce strong features on some of the observed properties of cosmic rays at low energies, such as the peak in the secondary-to-primary ratios at $\sim 1$ GeV/nucleon. However, earlier studies have shown that this kind of re-acceleration is not sufficient enough to produce any noticeable features in the cosmic-ray spectrum at high energies \cite{bib:Seo1994}. The efficiency of distributed re-acceleration decreases with energy, and its effect becomes negligible at energies above $\sim 20$ GeV/nucleon. On the other hand, for the re-acceleration by old supernova remnants shock waves discussed in this paper, the  efficiency does not depend strongly on energy. It depends primarily on the rate of supernova explosions in the Galaxy and the volume occupied by the supernova remnants. Its effect can be larger at higher energies as compared to the effect of  distributed re-acceleration.       

\section{Cosmic-ray transport equation with re-acceleration}
The re-acceleration of cosmic rays by weak shocks in the Galaxy is incorporated in the cosmic-ray transport equation as an additional source term following a power-law momentum spectrum \cite{bib:Wandel1987}. The cosmic-ray transport equation including diffusion, re-acceleration and interaction losses then follows,
\begin{equation}
\nabla\cdot(D\nabla N)-\left[\left(\bar{n} v\sigma+\xi\right)N+\xi sp^{-s}\int^p_{p_0}du\;N(u)u^{s-1}\right]\delta(z)=-Q\delta(z)
\end{equation}
where a cylindrical spatial coordinate $(r,z)$ system is adopted with the origin located at the center of the Galaxy. $N(\textbf{r},p)$ represents the differential number density of nuclei with momentum/nucleon between $p$ and $p+dp$, $D(p)$ is the diffusion coefficient, and $Q(r,p)$ represents the source term. The first term on the left-hand side of Eq. (1) represents diffusion. The second and third terms represent losses due to the inelastic interaction of cosmic rayss with the interstellar matter and due to the re-acceleration to higher energies respectively, where $\bar{n}$ represents the averaged surface density of interstellar atoms, $v(p)$ the particle velocity, $\sigma(p)$ the inelastic collision cross-section, and $\xi$ corresponds to the rate of re-acceleration. The fourth term containing an integral represents the population of particles produced by the re-acceleration of low-energy particles, assuming that particles are instantaneously re-accelerated to generate a power-law distribution of index $s$. The effects of ionization losses and convection due to the Galactic wind are neglected as these effects are mostly important at energies below $\sim 1$ GeV/nucleon. Our calculation concentrates only at energies above $1$ GeV/nucleon.

The propagation region is assumed to be bounded at $z=\pm H$, and unbounded in the $r$ direction. The matter and the sources are assumed to follow a uniform distribution in the Galactic disk with a radial size $R$. For cosmic-ray primaries, the source term in Eq. (1) is taken as $Q(r,p)=\bar{\nu} \mathrm{H}[R-r]\mathrm{H}[p-p_0]Q(p)$, where $\bar{\nu}$ represents the supernova explosion (SNe) rate per unit surface area of the disk, $\mathrm{H}(m)=1 (0)$ for $m>0 (<0)$ is the Heaviside step function, and $p_0$ is the lower cut-off in momentum/nucleon that has been introduced to account for the losses due to the ionization process. The source spectrum is assumed to follow a power-law in total momentum with an exponential cut-off at high energy. It can be written as function of momentum/nucleon as, 
\begin{equation}
Q(p)=AQ_0 (Ap)^{-q}\exp\left(-\frac{Ap}{Zp_c}\right)
\end{equation}
where $A$ and $Z$ respectively represents the mass and charge numbers of the nuclei, $Q_0$ is a constant corresponds to the fraction $f$ of kinetic energy of a supernova explosion that is injected into a given cosmic-ray species, $q$ represents the source spectral index, and $p_c$ is the cut-off momentum for protons at high energies. We have assumed that the maximum total momentum for a cosmic-ray nuclei is $Z$ times that of the protons. The cosmic-ray diffusion coefficient is assumed to follow $D(\rho)=D_0\beta(\rho/\rho_0)^a$, where $\rho=Apc/Ze$ is the particle rigidity with $e$ representing the charge of an electron and $c$ the velocity of light, $D_0$ is the diffusion constant, $a$ is the diffusion index, $\beta=v/c$, and $\rho_0$ is a constant. In Eq. (1), the re-acceleration parameter is written as $\xi=\eta V\bar{\nu}$, where $V=4\pi \Re^3/3$ is the volume occupied by a supernova remnant of radius $\Re$, and $\eta$ is a parameter introduced to take care of the unknown actual volume of supernova remnants that re-accelerate cosmic rays. We take $\Re=100$ pc which is approximately the size of a supernova remnant of age $10^5$ yr expanding in the interstellar medium with an initial shock velocity of $10^9$ cm s$^{-1}$. 

Eq. (1) is solved using the standard Green's function technique. The solution at $r=0$ is given by,
\begin{equation}
N(z,p)=\bar{\nu} R\int^{\infty}_0 dk\; \frac{\sinh\left[k(H-z)\right]}{\sinh(kH)}\times \frac{\mathrm{J_1}(kR)}{L(p)}
\times F(p) 
\end{equation}
where $\mathrm{J_1}$ is a Bessel function of order 1, 
\begin{equation}
L(p)=2D(p)k\coth(kH)+\bar{n}v(p)\sigma(p)+\xi,
\end{equation}
\begin{equation}
F(p)=Q(p)+\xi sp^{-s}\int^p_{p_0}u^s du\;Q(u)A(u)\times\exp\left(\xi s\int^p_u A(w)dw\right) 
\end{equation}
and, the function $A$ is given by
\begin{equation}
A(x)=\frac{1}{xL(x)}
\end{equation}

The local cosmic-ray density is obtained by taking $z=0$ in Eq. (3). The first term on the right-hand side of Eq. (5) represents  the normal cosmic-ray component, and the second term represents the re-accelerated component. Re-acceleration takes away particles from the low-energy part of the spectrum and puts them into the higher energy region. Therefore, for re-acceleration produced by weak shocks where $s>q$, the re-accelerated component might produce visible signatures, such as a bump or enhancement, in the energy spectrum. For re-acceleration by strong shocks which produces a harder particle spectrum, for instance $s=q$, the re-acceleration effect might not become visible since both the components follow the same spectra  \cite{bib:Wandel1987}.  

The spectrum for cosmic-ray secondaries in the Galaxy $N_2(\textbf{r},p)$ is obtained by solving a similar transport equation to their primaries with the source term given by,
\begin{equation}
Q_2(\textbf{r},p)=\bar{n} v_1(p)\sigma_{12}(p)H[R-r]H[p-p_0]N_1(\textbf{r},p) \delta(z)
\end{equation}
where $v_1$ represents the velocity of the primary nuclei, $\sigma_{12}$ represents the total fragmentation cross-section of the primary to the secondary, and $N_1$ is the density of the primary nuclei given by Eq. (3). The subscripts $1$ and $2$  denote the primary and secondary nuclei respectively.

The secondary-to-primary ratio is calculated by taking the ratio $N_2/N_1$. For no re-acceleration, i.e. $\xi=0$, it can be checked that Eq. (3) reduces to the standard solution of pure-diffusion equation, and the secondary-to-primary ratio becomes inversely proportional to the diffusion coefficient at high energies (see e.g., \cite{bib:Thoudam2008}). It can be mentioned that the energy spectrum of the secondary nuclei is more sensitive to the effect of re-acceleration than the spectrum of the primary nuclei \cite{bib:Thoudam2014}. Consequently, the secondary-to-primary ratios will also be more sensitive to the re-acceleration effect than the primary energy spectrum. For a steeper re-acceleration index $s>q$, the ratio is expected to exhibit an enhancement at lower energies, and for a harder index $s=q$, the ratio will have a flattening at high energies \cite{bib:Berezhko2003, bib:Wandel1987}. For the present calculation, since we are considering re-acceleration mainly by old supernova remnants, we will only consider the case of $s>q$ with $s\gtrsim 4$, which corresponds to a shock  Mach number of $\sim 1.7$.

\section{Results and discussions}
\begin{figure}[t]
\centering
\includegraphics[width=0.45\textwidth, angle=-90]{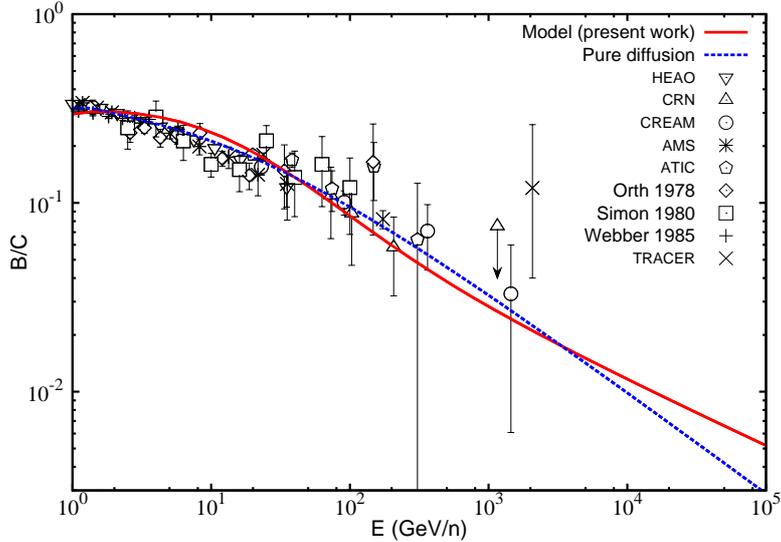}
\caption{Boron-to-Carbon (B/C) ratio. \textit{Solid line}: Present work including re-acceleration. \textit{Dashed line}: Pure diffusion model without re-acceleration \cite{bib:Thoudam2013}.}
\end{figure}
We take $H=5$ kpc, and the averaged surface density of the atomic hydrogen as $\bar{n}=7.24\times 10^{20}$ atoms cm$^{-2}$ \cite{bib:Thoudam2013}. We assume that the interstellar medium consists of $10\%$ helium. The inelastic interaction cross-sections are taken to be the same used in the calculations in Ref. \cite{bib:Thoudam2013}. We take the size of the source distribution $R=20$ kpc, and the proton low and high-momentum cut-offs as $p_0=100$ MeV/c and $p_c=1$ PeV/c respectively. The supernova explosion rate is taken as $\bar{\nu}=25$ SNe Myr$^{-1}$ kpc$^{-1}$, which corresponds to $\sim 3$ SNe per century in the Galaxy. The followings are taken as model parameters: the cosmic-ray propagation parameters $(D_0, \rho_0, a)$, the re-acceleration parameters $(\eta, s)$ and the source parameters $(q, f)$.
\begin{figure}[t]
\centering
\includegraphics[width=0.45\textwidth, angle=-90]{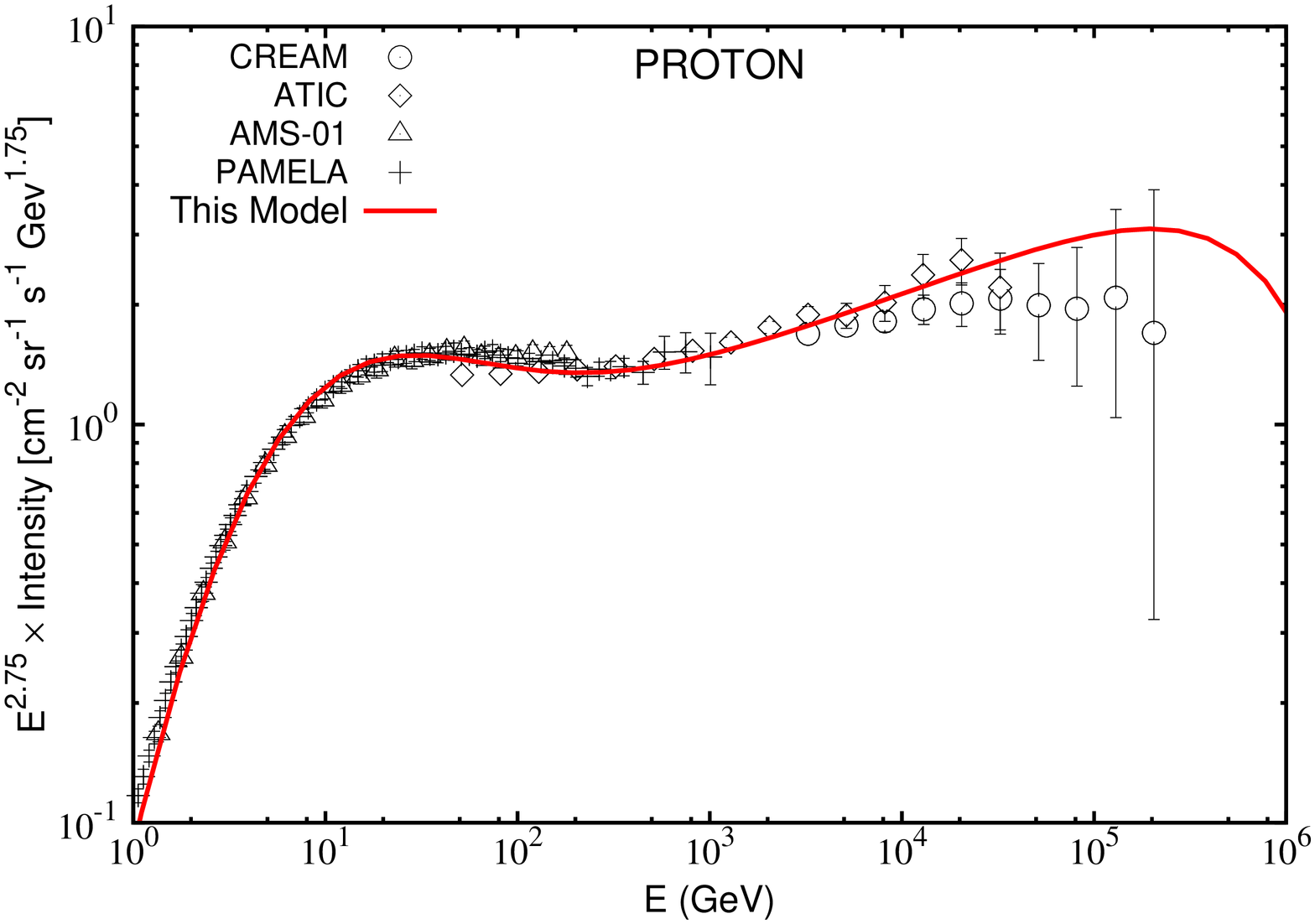}\\
\includegraphics[width=0.45\textwidth, angle=-90]{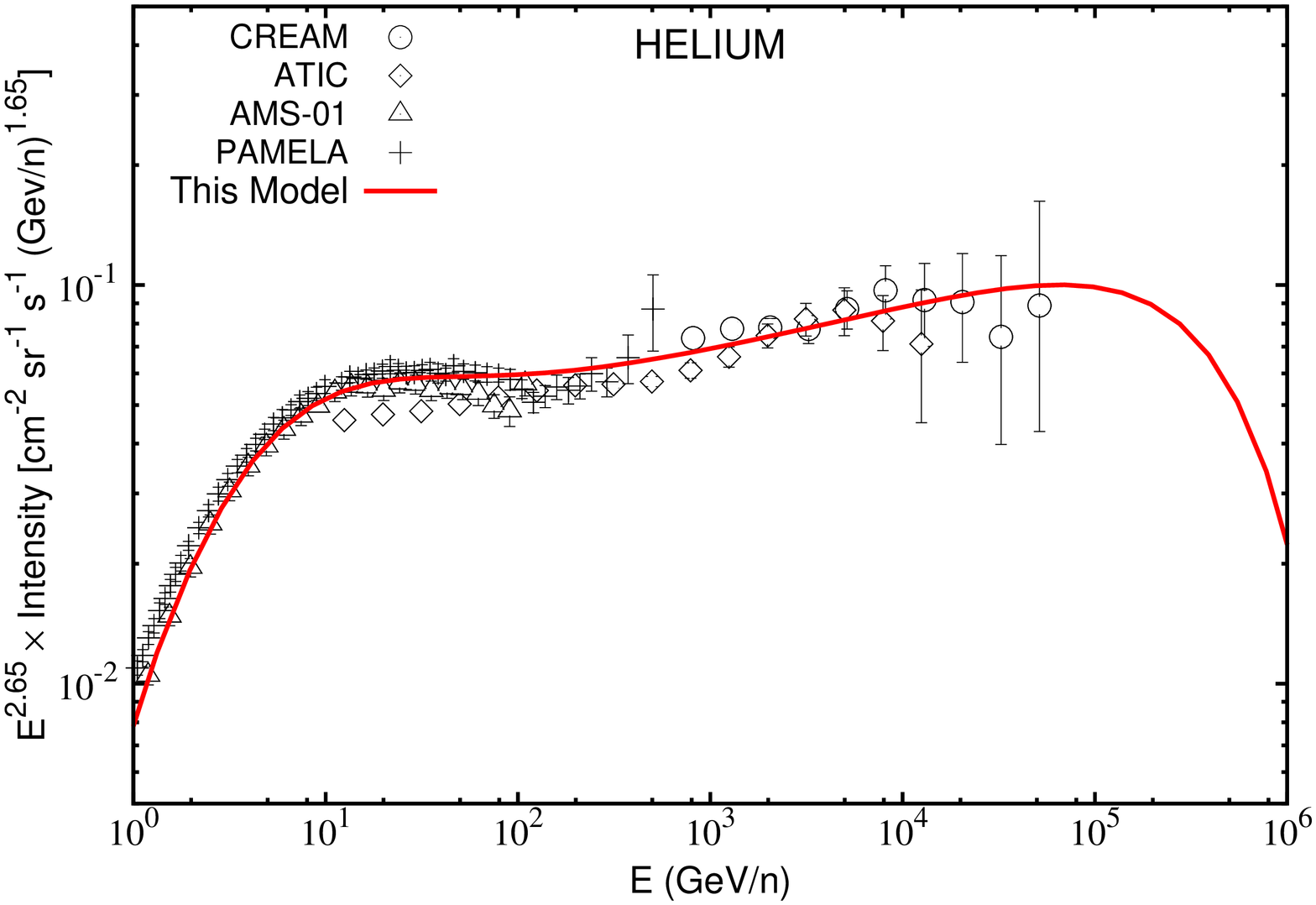}
\caption{\textit{Top}: Proton spectrum. \textit{Bottom}: Helium spectrum. The lines represent our results. For the data, see the experiments listed in Ref. \cite{bib:Thoudam2013}.}
\end{figure}

The values of $(D_0, \rho_0, a, \eta, s)$ are determined based on the measurements of boron-to-carbon ratio, and the measured energy spectra of carbon, oxygen, and boron nuclei. The values that gives the over-all best agreement between the model calculation and the measurements are found to be $D_0=9\times 10^{28}$ cm$^2$ s$^{-1}$, $\rho=3$ GV, $a=0.33$, $\eta=1.02$, $s=4.5$. These values correspond to the maximum amount of re-acceleration permitted by the available boron-to-carbon data. The result on the boron-to-carbon is shown in Figure 1 (solid line), where we have also shown the result for the case of pure diffusion (dashed line) with no re-acceleration $(\eta=0)$ taken from Ref. \cite{bib:Thoudam2013}. The best-fit carbon and oxygen source parameters are found to be $q_C=2.24, f_C=0.024\%$, and $q_O=2.26$, $f_O=0.025\%$ respectively, where the $f$'s are given in units of $10^{51}$ ergs. The calculation assumes a force-field solar modulation parameter of $\phi=450$ MV.

The spectra for the protons and helium nuclei calculated using the best-fit values of $(D_0, \rho_0, a, \eta, s)$ are shown in Figure 2. The top panel represents proton and the bottom panel represents helium nuclei. The data are the same as used in Ref. \cite{bib:Thoudam2013}. The source parameters required to explain the measurements are found to be $q_p=2.21, f_p=6.95\%$ for protons, and $q_{He}=2.18, f_{He}=0.79\%$ for helium. It can be seen that the model reproduces the measured data quite well, explaining the observed spectral anomaly between the GeV and TeV energy regions. 

\begin{figure}[t]
\centering
\includegraphics[width=0.45\textwidth, angle=-90]{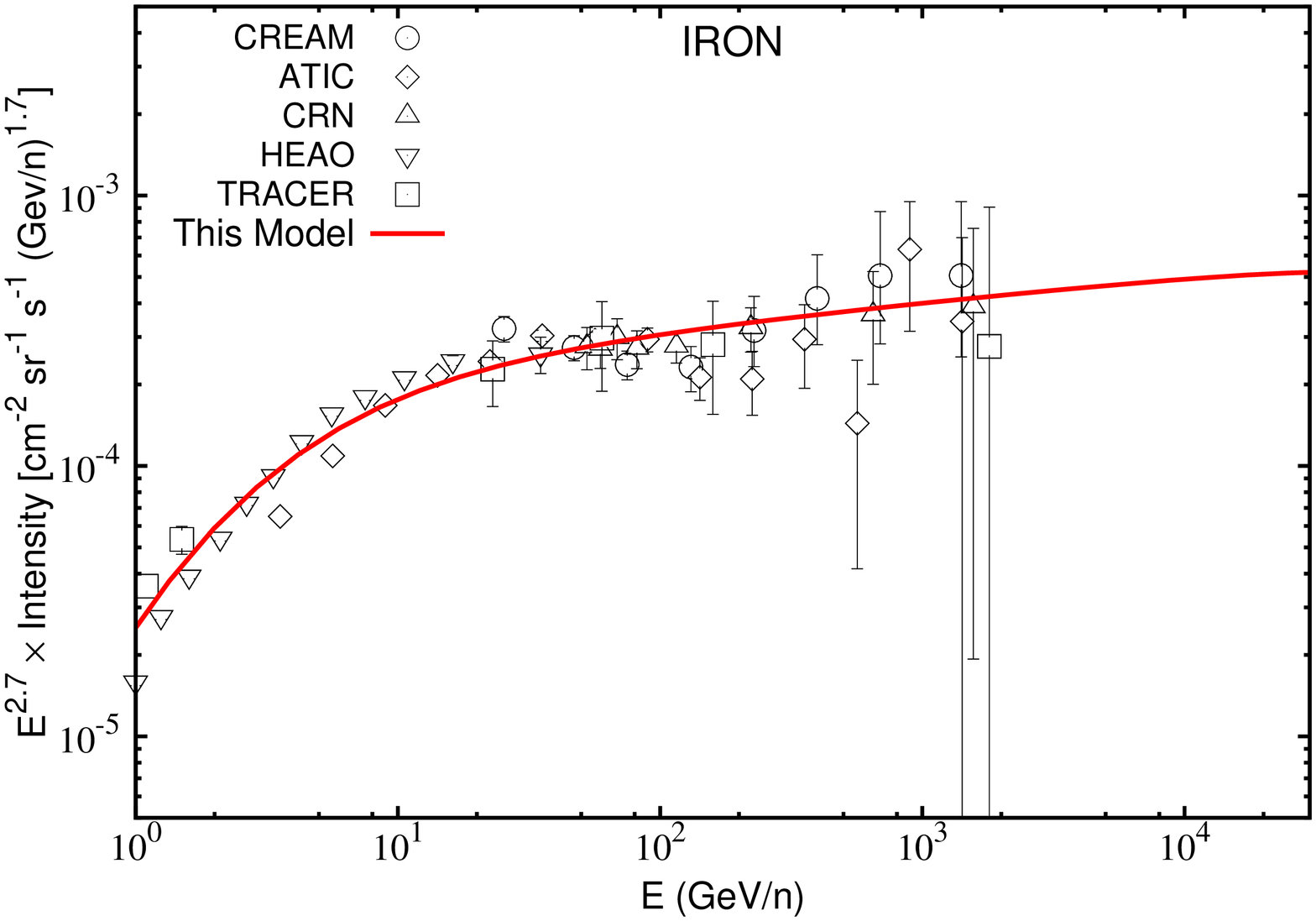}
\caption{Iron spectrum. The line represents our result. For the data, see the experiments listed in Ref. \cite{bib:Thoudam2013}.}
\end{figure}

In Figure 2, the effect of re-acceleration is stronger for protons than for helium nuclei. This is partly due to the steeper source spectrum for protons, as compared to the helium nuclei, required to reproduce the overall observed spectrum. For a steeper source spectrum, a larger fraction of low-energy particles becomes available for re-acceleleration leading into an increase in the re-accelerated component relative to the normal cosmic-ray component. But, the main reason for the stronger re-acceleration effect in the case of protons is due to their smaller inelastic collision cross-sections as compared to the helium nuclei. This means that low-energy protons can be efficiently re-accelerated to higher energies before they undergo inelastic collision with the interstellar matter during their residence time in the Galaxy. For helium nuclei, the loss due to the inelastic collision becomes important relative to the re-acceleration. For heavier nuclei such as iron for which the inelastic collision cross-sections are even larger, the collision losses becomes more important and the re-acceleration effect is expected to be negligible. This is shown in Figure 3 where we have compared our model prediction for the iron nuclei with the measurements. The calculation assumes $q_{Fe}=2.28$, and $f_{Fe}=4.9\times 10^{-3}\%$. As expected, the re-acceleration effect is hard to notice in Figure 3, and the model prediction above $\sim 20$ GeV/n follows approximately a single power-law unlike the proton and helium spectra. 

\section{Conclusions}
We have presented a model in which cosmic rays, after acceleration by strong supernova remnant shock waves, are re-accelerated by weak shocks associated with old supernova remnants during propagation in the Galaxy. The re-acceleration produces a cosmic-ray component that dominates the low energy part of the spectrum, while keeping the high-energy part dominated by the population that have not suffered re-acceleration. The model is found to explain the cosmic-ray spectral anomaly at GeV-TeV energies recently observed for the protons and helium nuclei. The study shows that the re-acceleration effect is important only for light nuclei, and is negligible for heavy nuclei like iron. Our prediction can be tested by sensitive measurements of heavy cosmic-ray species in future.

\end{document}